# Cluster Fragments in Amorphous Phosphorus and their Evolution under Pressure

Yuxing Zhou,[1] William Kirkpatrick,[1] and Volker L. Deringer[1]*

[1]*Department of Chemistry, Inorganic Chemistry Laboratory, University of Oxford, Oxford OX1 3QR, UK*

*E-mail: volker.deringer@chem.ox.ac.uk

## Abstract

Amorphous phosphorus (a-P) has long attracted interest because of its complex atomic structure, and more recently as an anode material for batteries. However, accurately describing and understanding a-P at the atomistic level remains a challenge. Here we show that large-scale molecular-dynamics simulations, enabled by a machine learning (ML)-based interatomic potential for phosphorus, can give new insights into the atomic structure of a-P and how this structure changes under pressure. The structural model so obtained contains abundant five-membered rings, as well as more complex seven- and eight-atom clusters. Changes in the simulated first sharp diffraction peak during compression and decompression indicate a hysteresis in the recovery of medium-range order. An analysis of cluster fragments, large rings, and voids suggests that moderate pressure (up to about 5 GPa) does not break the connectivity of clusters, but higher pressure does. Our work provides a starting point for further computational studies of the structure and properties of a-P, and more generally it exemplifies how ML-driven modeling can accelerate the understanding of disordered functional materials.

**Introduction**

Machine learning (ML) approaches are having a profound impact on computational modeling and data analysis across the various fields of materials science.[1-3] One of the emerging directions is the use of ML algorithms to create fast, yet powerful interatomic potential models for atomistic simulations.[4-6] "Machine-learned" potentials, or force fields, are fitted to a suitably chosen reference set of quantum-mechanical data, and so create a mathematical model of the potential-energy surface that describes the interactions between atoms.[7-11] Once an ML potential has been developed and carefully validated, it enables simulations that are several orders of magnitude faster than those with established quantum-mechanically based methods, such as "*ab initio*" molecular dynamics (AIMD). ML-based interatomic potentials, therefore, are beginning to be applied to a range of challenging materials-science research questions, such as the modeling of phase-change memory materials,[12-14] catalysts,[15] or battery materials.[16]

Recently, a number of "general-purpose" ML potentials have been reported, which can accurately describe a broad range of atomic configurations and materials properties – including silicon,[17] carbon,[18] aluminum,[19, 20] and the binary Ga–As system.[21] The hope for such potentials is to enable "off-the-shelf" use without further modification: for example, the aforementioned silicon ML potential has been used to study complex structural transitions under pressure[22] or unusual mechanical properties of amorphous silicon (a-Si).[23] The starting point for the present study is a general-purpose Gaussian approximation potential (GAP) ML model for bulk and nanostructured phosphorus, which was shown to be applicable to the pressure-induced liquid–liquid phase transition from the molecular fluid to the network liquid.[24] This GAP is now set to facilitate even more challenging studies on more extended length or time scales, and the exploration of other structurally complex phases for which it has not been explicitly "trained", such as amorphous phosphorus (a-P).

Research interest in a-P has grown because of emerging applications in batteries.[25-28] As a commercially available anode material, red phosphorus provides a large cation-storage ability with high theoretical capacities by forming binary X–P compounds (X = Li, Na, K), but it suffers from low conductivity and a large volumetric change during cycling.[29] As discussed in Ref. [30], these issues can be ameliorated by creating composites of a-P and carbonaceous materials: on the one hand, increasing electronic conductivity;[31, 32] on the other hand, minimizing the mechanical stress induced by volume changes.[30, 33] The atomistic structure of phosphorus itself is clearly important here: for example, experimental work showed that the size of a-P particles in phosphorus–carbon composite anodes has an effect on the electrochemical performance.[34] A closer understanding of the local structure of the pristine material, including its response to pressure (which is expected to change locally during battery operation), would therefore be highly beneficial. If successful, one might hope that atomic-scale simulation studies of the structure of a-P, and its evolution under pressure, could ultimately help to understand and control the electrochemical properties of a-P based anodes.

What, exactly, is the structure of a-P? Various models have been proposed to date. Lannin *et al.* suggested a quasi-two-dimensional structure with locally layer-like fragments, resembling those in black phosphorus, supported by Raman scattering and infrared absorption.[35-37] Elliott and coworkers argued that neutron diffraction data support the presence of $P_8$ (and possibly $P_9$) cage-like fragments;[38] hence, the medium-range order (MRO) in a-P might be similar to that found in crystalline Hittorf's (violet) and fibrous phosphorus.[39, 40] This interpretation is consistent with early Raman spectroscopy studies, which also identified the presence of $P_8$ and $P_9$ cages[41, 42] and likely formation of $P_7$ clusters.[42] By contrast, Zaug *et al.* carried out empirical potential structure refinement (EPSR) based on X-ray diffraction (XRD) data,[43] and the resulting structural model mainly comprises $P_3$ rings and $P_4$ tetrahedra linked by chains of atoms. These $P_3$ and $P_4$ units are considerably different from the aforementioned $P_7$, $P_8$, and $P_9$ ones, for which the principal structural unit is a 5-membered ring.

In addition to experimental studies, computational work provided insights into possible cluster fragments in the amorphous phase. Early AIMD simulations of $P_2$–$P_8$ clusters and of melt-quenched a-P led Jones and Hohl to suggest a network structure of small *n*-membered clusters.[44, 45] A systematic quantum-mechanical study of structural fragments was reported by Böcker and Häser,[46] building on Baudler's rules for describing phosphorus structures.[47] The authors suggested that the dominant microscopic structures in a-P are tubular units and chains, consisting of clusters containing 7–10 atoms that had been discussed earlier.[47, 48] Very recently, simulated Raman spectra for crystalline allotropes and other candidate structures corroborated the absence of small chain fragments in a-P, including $P_4$ tetrahedra.[49] Yet, all these studies are invariably limited by the system sizes that are accessible to quantum-mechanical methods.

In the present work, we report large-scale ML-based molecular-dynamics simulations on a timescale of nanoseconds (corresponding to millions of individual simulation steps) to study structural features and pressure-dependent changes of a-P. In doing so, we expand substantially on previous ML-driven studies of the structure of hypothetical crystalline allotropes[50, 51] and the liquid structures of phosphorus[24] – here reporting the first application of ML potentials to the amorphous phase of the element, to our best knowledge. We show that the structural model of a-P generated in this work, by simulated quenching from a disordered melt, agrees with existing experimental data in terms of calculated structure factors. We then study changes in the MRO of a-P during simulated compression and decompression, and how these changes can be linked to the diffraction "fingerprint" of the material. An intrinsic connection among cluster fragments, rings, and voids is revealed, which may help with the interpretation of existing and perhaps future experimental data. Our work provides an example of how ML-based interatomic potentials can be used in the study of a-P, as well as other complex disordered functional materials that are of interest for practical applications.

## Results and Discussion

To create structural models of a-P, we carried out melt-quench simulations (**Figure 1**). The high-temperature liquid at ambient pressure consists mainly of $P_4$ molecules, whereas it transforms to a network liquid at higher pressure.[52] We use the latter as a starting point, decompressing it, and then cooling from that metastable liquid state: the protocol is indicated in Figure 1a. We used a quench rate of $10^{11}$ K s$^{-1}$ from 1200 K to 300 K (details are given in the Methods section) – more than three orders of magnitude slower than in a previous AIMD simulation of a-P,[45] and consistent with recent GAP-driven studies of a-Si.[53]

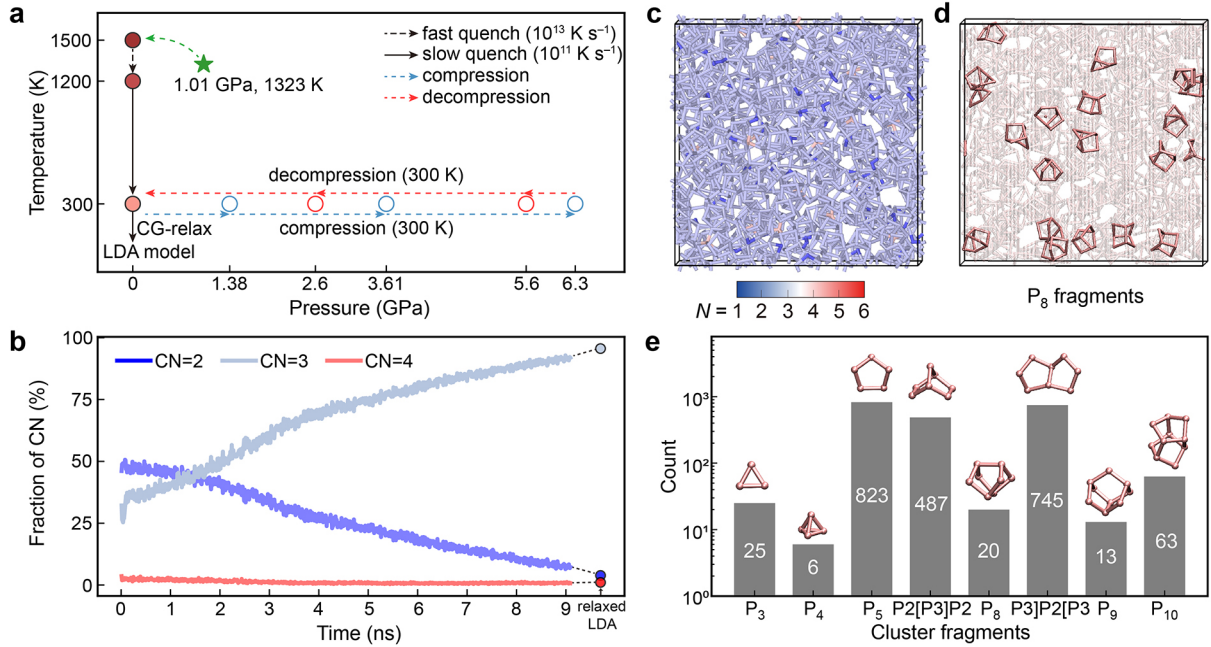

**Figure 1.** The structure of amorphous phosphorus as described by machine-learning-driven simulations. (**a**) An overview of the protocols used in this work to create the low-density amorphous (LDA) and high-density structural models. Starting from a disordered network liquid, we carry out a fast quench ($10^{13}$ K s$^{-1}$) to 1200 K, and then a much slower quench ($10^{11}$ K s$^{-1}$) from 1200 to 300 K. A conjugate-gradient (CG) relaxation is performed for the amorphous structural model before compression and subsequent decompression at 300 K. (**b**) Coordination numbers (CN) during the quench, based on the count of nearest neighbors with a bond-length cutoff of 2.4 Å. The relaxed LDA structure has 4% two-fold, 95% three-fold, and 1% four-fold coordinated atoms, as indicated by symbols (dotted lines are guides to the eye). (**c**) Atomic structure of the LDA model, relaxed after the end of the slow quench down to 300 K. The CN is indicated by color coding. (**d**) Highlighted $P_8$ cage-like fragments in the simulated LDA model. (**e**) Counts of relevant cluster fragments as found in the LDA model. Exemplary images of the respective clusters are shown.

We emphasize that the protocol shown in Figure 1a does not mirror experimental conditions directly – but it allows the simulation to probe a variety of possible fragments, on account of the large disorder in the network liquid, and thereby to find local energetic minima. This way, we obtain a candidate structural model which may be (and needs to be!) validated against experiment. Following the quench to 300 K, we fully relaxed the structure, and refer to this as "low-density amorphous" (LDA) in the following, borrowing a term that is often used in the field of disordered silicon.[54]

Figure 1b characterizes the changes in coordination number (CN) statistics during the quench: there is a gradual decrease in the number of two-coordinated atoms (CN = 2), and a steady increase for CN = 3. The latter is expected for solid phosphorus as a fifth-main-group element, and found in the ambient-pressure crystalline allotropes throughout. Very few four-coordinated atoms (CN = 4), which we regard as defective, over-coordinated environments, exist in our structural model; most of them are eliminated during the quench process. In the final, relaxed structure, about 95 % of the atoms have CN = 3 (Figure 1b–c).

To analyze the cluster fragments, which consist of different combinations of primitive rings, we started from the results of shortest-path rings statistics and counted the numbers of the rings and more complex cluster fragments. For example, the occurrences of the $P_8$ cage-like fragment, which consists of four fused five-membered rings, are highlighted in Figure 1d; a more comprehensive overview is provided in Figure 1e. The dominant building units in our LDA model are primitive 5-membered rings, labeled "$P_5$". Most of these rings are furthermore involved in the formation of multiple more complex clusters, including the seven-atom "P2[P3]P2" fragment (two fused 5-membered rings with 3 shared atoms), the eight-atom "P3]P2[P3" one (two fused 5-membered rings with 2 shared atoms), and $P_{10}$ cages. These fragments, and the way that complex fragments are related to the constituent simpler ones, have been discussed in detail by Böcker and Häser.[46] Only 12.9% of the atoms in the simulation cell (256 out of 1,984) take part in the formation of $P_8$ / $P_9$ cages, much fewer than in crystalline Hittorf's (violet) and fibrous phosphorus (81%).[39, 40] This is in line with earlier experimental work[42] in which it was stated that $P_8$ cages were present in much reduced numbers in a-P.

Besides, our LDA structure only contains few $P_3$ rings and $P_4$ tetrahedra, 25 and 6 respectively, at variance with the model discussed in Ref. [43]. Hence, our structural model, generated with a potential "machine-learned" from quantum-mechanical data, has features similar to those proposed by Böcker and Häser on the basis of gas-phase computations.[46] The low abundance of $P_4$ tetrahedra is consistent with reported Raman spectra, in which the intensity of the peak representing $P_4$ was much lower than for those indicating the presence of the $P_7$, $P_8$, and $P_9$ fragments.[42, 55, 56]

The calculated mass density of our a-P structural model is 2.25 g cm$^{-3}$, consistent with the reported experimental data ranging from 2.14 to 2.34 g cm$^{-3}$.[43, 57] Notably, this density is lower than that of Hittorf's (violet) phosphorus (2.36 g cm$^{-3}$),[39] as well as that of orthorhombic black phosphorus (2.71 g cm$^{-3}$),[58] indicating a more open network – consistent with the appearance of the structure (Figure 1c–d), which contains covalently linked cluster fragments that are separated from further clusters by abundant voids. The GAP-computed energy of our structural model is 7 kJ mol$^{-1}$ (71 meV at.$^{-1}$) higher, *i.e.*, less favorable, than that of layered black phosphorus, but 8 kJ mol$^{-1}$ (84 meV at.$^{-1}$) lower than that of molecular white phosphorus, and slightly higher than earlier predictions for more ordered "nanotubular" allotropes which have been experimentally characterized[59] and studied using dispersion-corrected density-functional theory.[60] The computed energy of our structural model is therefore in line with what one would expect for a metastable, but experimentally readily accessible amorphous material.

Beyond the ambient structure of a-P, pressure-induced structural transitions in the material are of interest, and have been characterized using experimental techniques.[43, 61, 62] For example, high-resolution transmission electron microscopy (HRTEM) indicated that nanocrystalline nucleation sites formed at the early stage of compression, which are expected to be mainly driven by the considerable volume decrease.[63] Indeed, equation-of-state measurements also suggested that a-P undergoes a substantial volume decrease by 30% when compressed to 8 GPa: a sharp drop in volume was observed at 7.5 GPa.[56] Further (*in situ*) XRD and Raman spectroscopy experiments showed that, following nucleation under pressure, a-P crystallizes into black phosphorus at 7.0 to 8.0 GPa.[56, 63]

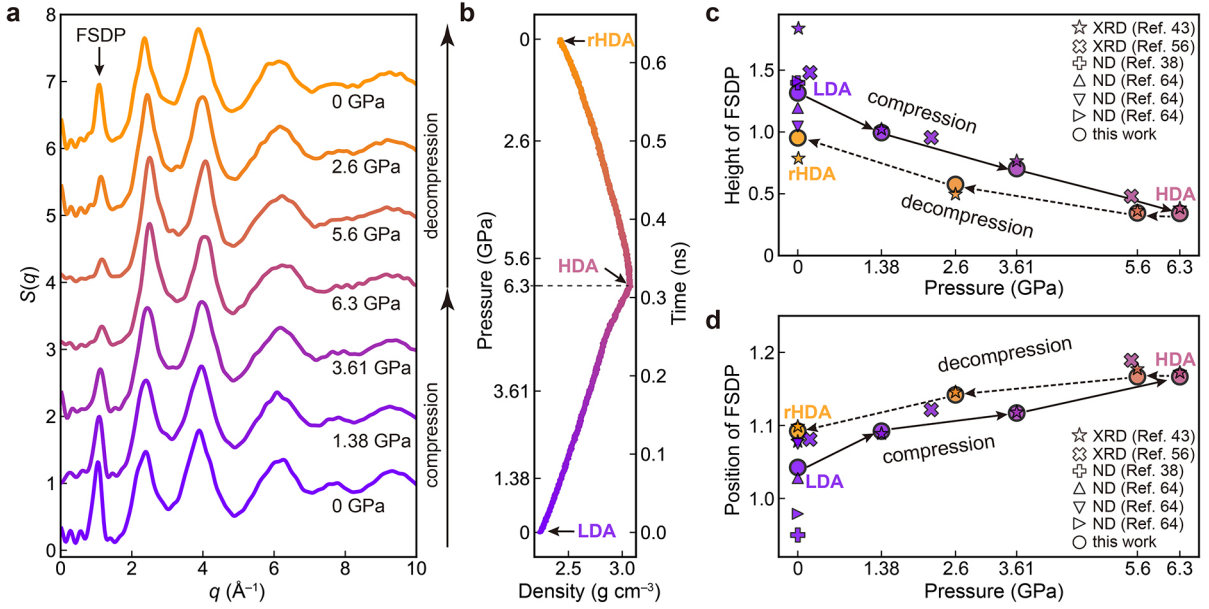

**Figure 2.** Pressure-dependent changes in the structure factor of amorphous phosphorus. (**a**) Calculated static structure factor, $S(q)$, at ambient temperature during the compression and decompression simulations. The structure factor was calculated based on the Fourier transformation of the computed radial distribution function, $g(r)$, with a real-space truncation distance of 25 Å. Data are vertically offset in steps of +1 for clarity. (**b**) Calculated mass density during compression and decompression. (**c–d**) Comparison of the heights and the positions of the FSDP between the calculated structure factors (Figure 2a) and experimental data taken from previous work.[38, 43, 56, 64] Results from different experiments, *i.e.*, X-ray diffraction (XRD) and neutron diffraction (ND) derived FSDP heights and positions, are shown by symbols.

To study the pressure-dependent behavior of the cluster fragments on the atomic scale, we started from our fully relaxed LDA structure and performed a compression and subsequent decompression simulation, at a temperature of 300 K. This simulation protocol (cf. Figure 1a) follows the experimental report in Ref. [43], albeit, necessarily, at a faster rate of compression. Our LDA structure is first compressed from ambient pressure up to 6.3 GPa, with a rate of 0.02 GPa ps$^{-1}$; the external pressure is then gradually released at the same rate until it reverts to ambient conditions. We call the compressed phase at 6.3 GPa the high density amorphous (HDA) phase, and the decompressed phase at ambient pressure "recovered HDA" (rHDA), to emphasize that it is different from LDA phosphorus as generated from the melt-quench process mentioned above – consistent with the high-pressure experimental study of Ref. [43]. The results of the compression–decompression simulations are characterized in **Figure 2**.

The first sharp diffraction peak (FSDP) in the structure factor, $S(q)$, may be affected by multiple aspects of the structure,[65, 66] including the distribution of voids as well as the MRO. The FSDP has also played an important role in understanding a-P.[43, 56] We calculated $S(q)$ for the configurations at the same pressure values as reported for the experimental compression–decompression cycle of Ref. [43] (Figure 2a), and additionally we evaluated the density change under pressure (Figure 2b). Our ML-driven simulations yield good agreement with earlier experiments in terms of the pressure-dependence of the FSDP:[43, 56] both the simulated intensities (heights) and the positions of the FSDP signals (Figure 2c and 2d, respectively) agree well with X-ray and neutron diffraction data. The largest discrepancy in this regard appears for our LDA model compared to ambient-pressure measurements, but these themselves show considerable variation (Figure S1),[38, 43, 56, 64] presumably due in parts to structural differences in various commercial samples. Indeed, a comparative neutron-diffraction study of two a-P samples synthesized at different temperatures revealed differences in the measured $S(q)$, which were interpreted in terms of changes in the intermediate-range order.[67] Besides, it was suggested that changes in the FSDP can be attributed to the average void size, void spacing, and void density,[43] which may differ between our simulated LDA structure and some experimental ones. Despite small deviations, the full ambient-pressure structure factor $S(q)$ computed in our work shows fair agreement with experimental data throughout (Figure S1).

During the compression and decompression cycle, we observed a hysteretic recovery, in which the changes in the FSDP only partly reversed upon decompression (Figure 2c–d) – leaving the atomic structure in a denser configuration (Figure 2b). The mass density of our recovered rHDA model after decompression is 2.42 g cm$^{-3}$, slightly higher than that of Hittorf's phosphorus (2.36 g cm$^{-3}$; Ref. [39]) but still lower than that of layered black phosphorus (2.71 g cm$^{-3}$; Ref. [58]). The right-shifted and less intense FSDP during decompression suggests a more compact and less ordered amorphous structure as compared to the initial one (LDA) undergoing compression, consistent with the density changes shown in Figure 2b and with those reported by Zaug *et al.*[43] We note that the sluggish recovery of structural features was also indirectly observed in Raman spectra for compression / decompression experiments.[56]

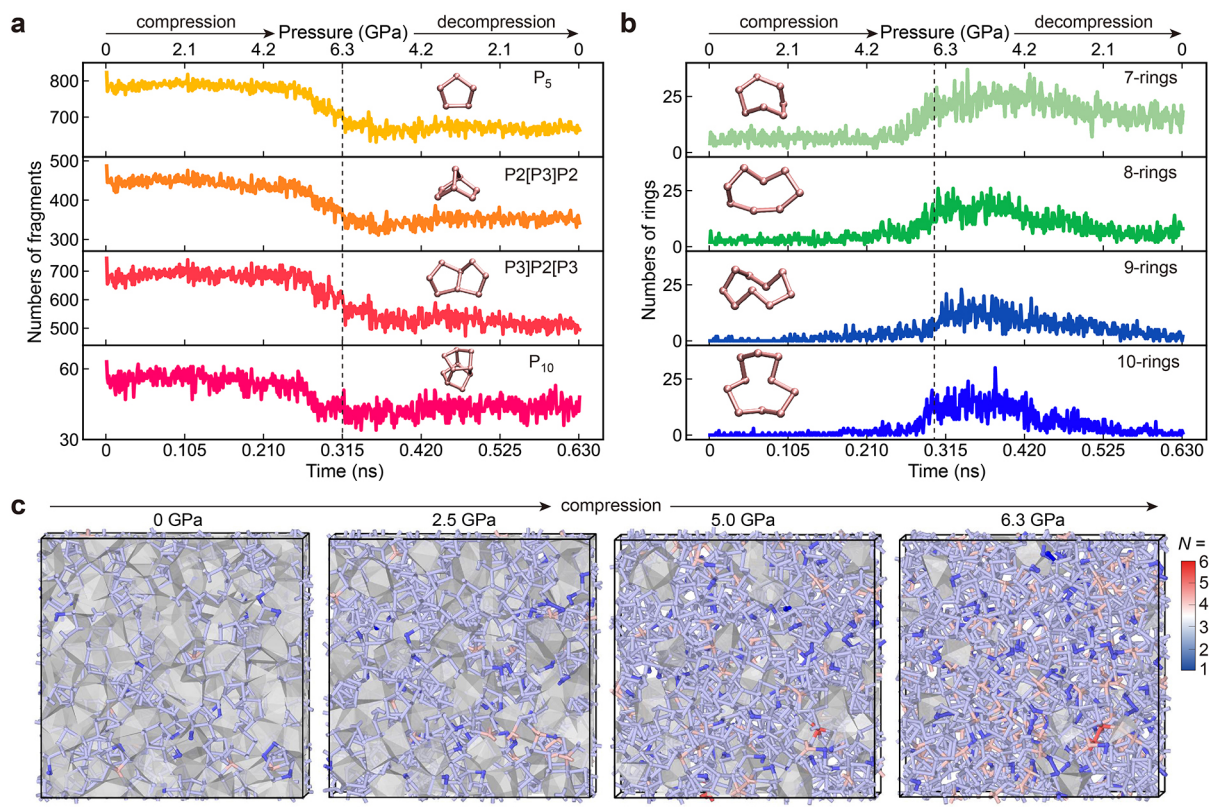

**Figure 3.** The evolution of atomic structure in amorphous phosphorus during simulated compression and decompression. (**a**) Count of selected cluster fragments, including $P_5$, P2[P3]P2, P3]P2[P3, and $P_{10}$, and (**b**) count of large rings, containing 7 to 10 atoms. Vertical dashed lines indicate the change from compression to decompression. Illustrative examples of clusters and rings are shown as insets. (**c**) Void analysis based on the alpha-shape method.[68] The radius of the probe sphere used here is 2.6 Å, slightly larger than the average nearest-neighbor distance in a-P (cf. Figure S3). The surface of the void region is rendered in gray.

To better understand the pressure-induced behavior of the cluster fragments, we analyzed the counts of clusters and higher-membered rings, as shown in **Figure 3**. At the start of the compression, up to about 5 GPa, the numbers of these cluster fragments fluctuate slightly but do not change markedly (Figure 3a), and almost no large rings with 7–10 members are found (Figure 3b), suggesting that modest pressure cannot immediately break the local structures. Upon further compression beyond 5 GPa, the number of cluster fragments decreases, which continues up to 6.3 GPa; meanwhile, high-membered rings start to form, which is attributed to the opening of cluster fragments. These observations, again, agree well with experiments: earlier *in situ* Raman experiments indicated that pressure up to 3 GPa does not strongly affect the local bonds, whereas all clear peaks found in the ambient Raman spectra become rather weak above 5 GPa, hinting at a significant deformation of the structural building units at higher pressure.[55] In addition, a very small peak at ≈ 600 cm$^{-1}$ was observed in the Raman spectrum of a-P,[55, 56] interpreted as the dominant Raman-active mode of $P_4$ molecules in white P,[69] and this peak disappeared under pressure.[55] We also found in our simulations that the count of remaining $P_3$ rings and $P_4$ tetrahedra starts to decrease at ≈ 4.5 GPa and then quickly disappears as the simulation reaches 6.3 GPa of external pressure (Figure S2).

In the subsequent decompression, from 6.3 GPa back to ambient conditions, those dominant cluster fragments did not immediately become more abundant but gradually re-appeared, further supporting the proposed distinction between LDA and rHDA forms of phosphorus (based on a partial retention of structural features as pressure is released). Meanwhile, distorted higher-membered rings are barely found at the start of the compression. The count of 7-membered rings increases sharply at about 4.5 GPa, whereas a similar increase takes place in 8-, 9- and 10-membered rings at ≈ 5 GPa (Figure 3b). In addition, these larger rings slowly disappear when pressure is gradually released, and many of them remain after decompression and in the rHDA phase, in contrast to what is found in LDA phosphorus.

A void analysis (see Methods section) was performed to better understand the evolution of the MRO in a-P, and the results are shown in Figure 3c. Upon compression to 5.0 GPa, we observed that most voids in LDA phosphorus are gradually squeezed out, and different cluster fragments are brought closer to each other yet remain intact, leading to the formation of relatively long P⋯P contacts (2.4–2.8 Å) between neighboring clusters. This is evidenced by the gradually increased magnitude of the radial distribution function around its first radial minimum during compression (Figure S3). As pressure further increases (from about 5.0 to 6.3 GPa), the clusters start to open – giving rise to the disappearance of LDA-like fragments (Figure 3a), and the concomitant increase in the count of larger rings (Figure 3b).

**Conclusions**

Computer simulations using a machine-learned interatomic potential have provided an atomic-scale picture of the local structure and the pressure-dependent structural changes in amorphous phosphorus. $P_5$ rings, P2[P3]P2 and P3]P2[P3 clusters, and $P_{10}$ cages are dominant building units in our structural model and are separated by abundant voids. $P_8$ and $P_9$ cages, the structural building units of Hittorf's and fibrous phosphorus, are found as well, whereas $P_3$ rings and $P_4$ tetrahedra are much less abundant. We performed a compression simulation on this structure, in which the external pressure squeezed out most of the voids at moderate pressure (up to 5 GPa). A continuous increase of pressure up to 6.3 GPa subsequently caused some of the cluster fragments to break, leading to the formation of larger rings. A hysteretic recovery of medium-range order was observed in terms of the heights and the positions of the FSDP. The agreement between simulated and (previously) experimentally obtained structure factors provides a firm basis for interpreting the simulation result, and it enables the study of atomic-scale structural changes that are not directly accessible to experimental observation.

Looking beyond the new findings regarding phosphorus, our work provides an example of the growing importance of ML-driven modeling for understanding structural changes in disordered functional materials. Such mechanisms are important both for fundamental insight and for practical applications in devices, and they are poised to become much better understood with the large-scale and quantum-mechanically accurate simulations that ML methods are increasingly making possible. Our studies have been performed with a general-purpose ML potential, and it is hoped that they provide an example of its usefulness – but also of the usefulness of ML potentials more generally; indeed, the liquid–liquid phase transition and phase diagram of phosphorus has recently been studied in more depth using an active-learning-based neural-network interatomic potential.[70] Hence, one may envision that ML potentials will, in the future, facilitate the exploration of the structural nature of a wide range of disordered solids, alongside synthesis and experimental characterization – and that they will provide us with further insight into how atomic structure and macroscopic properties are connected in amorphous functional materials.

## Methods

**ML-driven MD simulations:** All MD simulations were performed in the NPT ensemble, using LAMMPS,[71] with a Nosé–Hoover thermostat controlling temperature[72, 73] and a barostat controlling external pressure.[74] To obtain the LDA a-P structure, the starting configuration was taken from the high-density liquid at 1323 K at 1.01 GPa, containing 1,984 atoms in a cubic simulation cell with a density of 2.78 g cm$^{-3}$ (green star in Figure 1a). The structure was first held at 1500 K for 50 ps, so as to be strongly randomized, and then cooled to 1200 K at a rate of 10 K ps$^{-1}$. We note that slow cooling provides a better-relaxed structure, leading to a more ordered network containing fewer defects; hence we slowly cooled down the liquid from 1200 K to 300 K with a quench rate of 0.1 K ps$^{-1}$. This quench rate was also used in ML-based cooling simulations to obtain a well-relaxed structure of a-Si.[53] The time step in all simulations was 1 fs. The obtained amorphous structure was further optimized based on the conjugate-gradient method (at fixed cell shape and volume, thereby retaining the density from the last step of the MD simulation) to obtain our final LDA phosphorus model.

Energies and forces on atoms were computed with a general-purpose ML interatomic potential for phosphorus developed in previous work.[24] This potential was fitted using the Gaussian approximation potential (GAP) framework[9, 75] together with the smooth overlap of atomic positions (SOAP) structural descriptor[76] and a long-range "+R6" baseline for van der Waals interactions.[18] The model was fitted to a large reference database including different crystalline allotropes of phosphorus, layered (two-dimensional) phosphorus, liquid phosphorus (both network and molecular phases), and random configurations generated via a random structure searching (RSS) approach.[50, 77] The labels on the reference data include many-body dispersion corrections[78, 79] to accurately describe the van der Waals interactions. The latter are of high importance in crystalline allotropes of phosphorus,[60, 80] and by extension we expect them to be relevant for the amorphous phase as well.

We note that in the previous work, a liquid–liquid phase transition simulation was performed primarily as a means of validating the potential, and configurations corresponding to both the molecular and the network liquid form had been included in the reference database. Here, in contrast, we demonstrate that this GAP potential can describe other disordered phases (namely, amorphous phosphorus at ambient temperatures and modest pressures) which are *not* explicitly included in the training database – thereby providing evidence for good transferability across structural space, as has recently been seen for disordered silicon under pressure.[22]

**Notation for cluster fragments:** The notations of all clusters discussed in this work are taken from Böcker and Häser's work,[46] and similar notations were also used in other studies of phosphorus clusters.[44, 47] We use the standard $P_n$ notation for clusters consisting of *n* phosphorus atoms, and employ square brackets enclosing parts of more complex fragments (for example, "[P3]") to describe the relationship between "donors and acceptors" as discussed in earlier work.[46, 47] The fragments we selected to be studied here are those discussed by both theoretical and experimental work, as mentioned in the main text. The analysis of cluster fragments and their connectivity was carried out with the help of the R.I.N.G.S. software.[81] It can transform the atomic structure into a graph with nodes for atoms and links for bonds, and therefore, different sizes of the closed shortest-path rings can be counted. We note that the 3-, 4-, and 5-membered rings are the building units of the more complex cluster fragments studied in this work, and thus the cluster fragments can be counted based on the arrangements of the rings found by the R.I.N.G.S software.

**Void analysis:** The void regions in the computed structures (indicated by gray surfaces in Figure 3c) are identified based on the alpha-shape method[68] as implemented in the OVITO software.[82] This method is based on the construction of a surface separating filled spatial regions and empty regions, assuming that all atoms are points (*i.e.*, without a radius). We used a probe radius of 2.6 Å to detect the voids, slightly larger than the position of the first minimum in the calculated radial distribution function of a-P (Figure S3).

**Visualization:** The structure pictures in Figures 1c and 3c were generated using OVITO;[82] those in Figure 1d–e were generated using VMD.[83]

## Acknowledgements

We thank I. Capone, S. R. Elliott, M. Pasta, and M. Wilson for useful discussions. Y.Z. acknowledges support from a China Scholarship Council-University of Oxford scholarship. This work was performed using resources provided by the Cambridge Service for Data Driven Discovery (CSD3) operated by the University of Cambridge Research Computing Service (www.csd3.cam.ac.uk), provided by Dell EMC and Intel using Tier-2 funding from the Engineering and Physical Sciences Research Council (capital grant EP/P020259/1), and DiRAC funding from the Science and Technology Facilities Council (www.dirac.ac.uk). The authors would like to acknowledge the use of the University of Oxford Advanced Research Computing (ARC) facility in carrying out this work (http://dx.doi.org/10.5281/zenodo.22558).

## Data Availability Statement

The data that support the findings of this study will be made openly available in the Zenodo repository upon journal publication.

Supporting Information for

# Cluster Fragments in Amorphous Phosphorus and their Evolution under Pressure

Yuxing Zhou,[1] William Kirkpatrick,[1] and Volker L. Deringer[1]*

[1]*Department of Chemistry, Inorganic Chemistry Laboratory, University of Oxford, Oxford OX1 3QR, UK*

*E-mail: volker.deringer@chem.ox.ac.uk

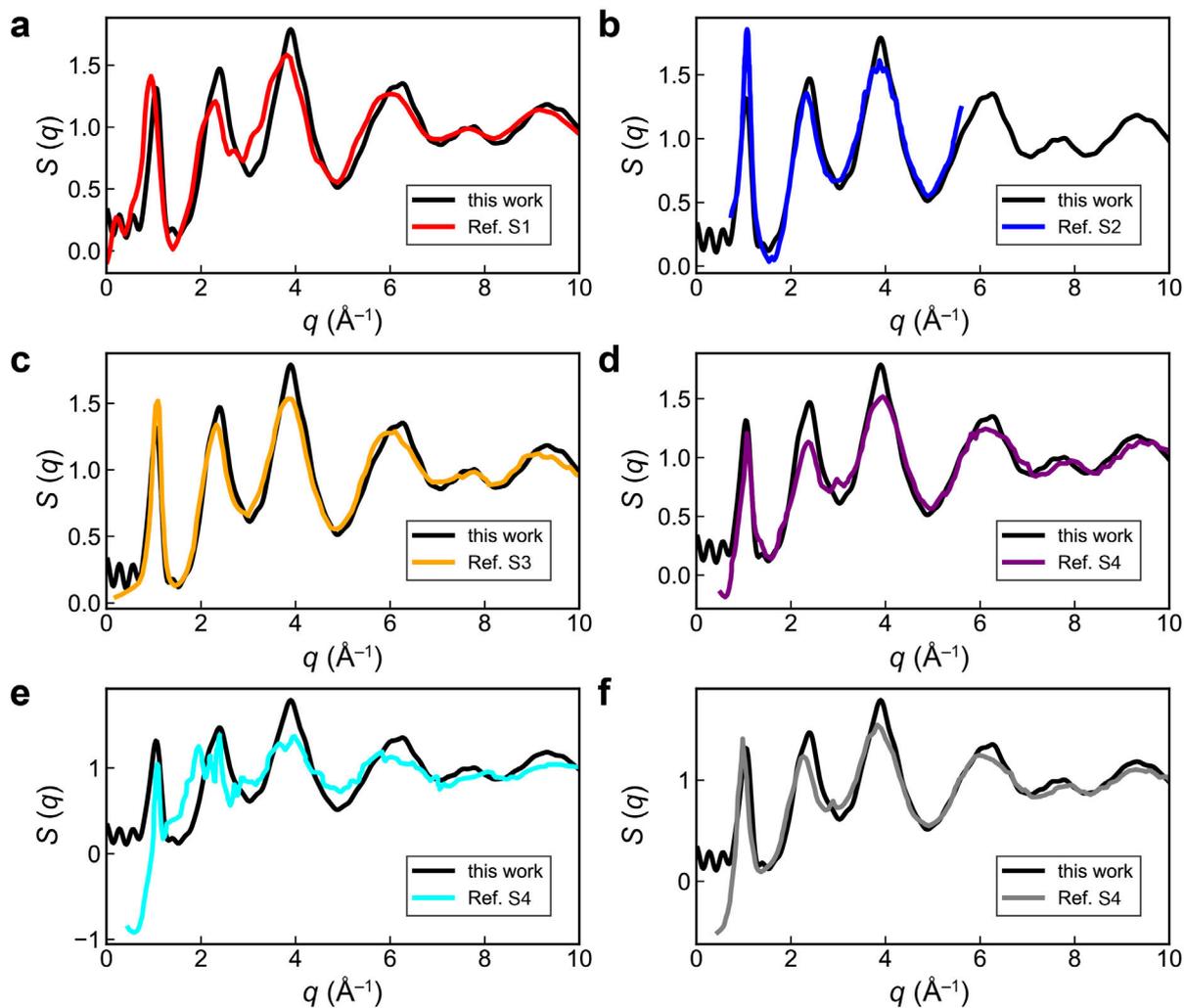

**Figure S1.** The computed ambient-pressure structure factor $S(q)$ for the structural model described in the present work (given in each panel as a black line), as compared to the experimental results given in (**a**) Ref. S1, (**b**) Ref. S2 (at 0.01 GPa), (**c**) Ref. S3 (at 0.2 GPa), and (**d**–**f**) Ref. S4 (where three different samples were investigated). These references are also cited as Refs. 38, 43, 56, and 64 in the main text, respectively. An important point emphasized by this figure is that a-P samples will depend on the synthesis protocol, and their structure (as reflected in the measured structure factor) varies from case to case.

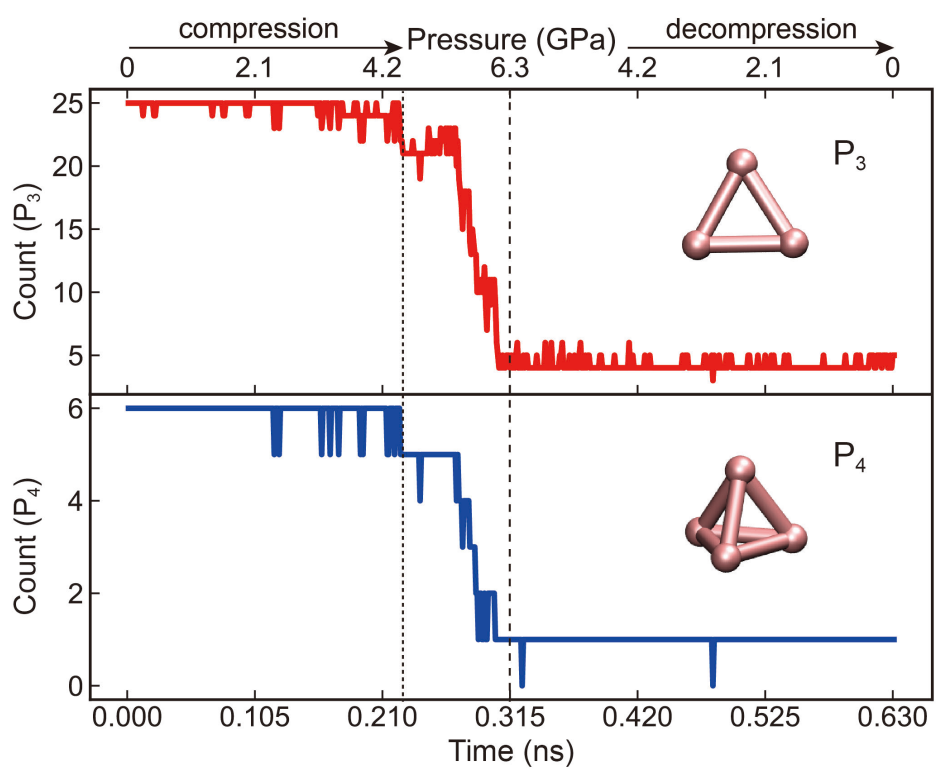

**Figure S2.** The evolution of the count of $P_3$ rings and $P_4$ pyramids during simulated compression of the LDA phosphorus structure. The dashed line in the middle represents the change from compression to decompression. Neither of these fragments is highly abundant in the initial structure, and a sharp decrease in the count of both $P_3$ and $P_4$ units occurs at about 4.5 GPa, as indicated by the dotted line.

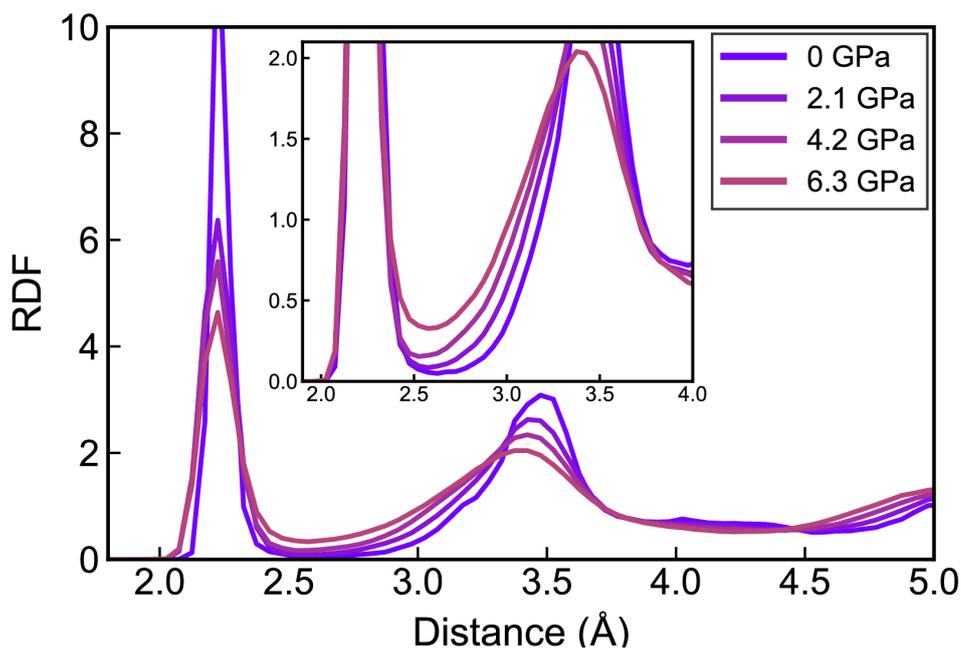

**Figure S3.** Calculated radial distribution functions (RDF) of a-P during the compression from 0 GPa to 6.3 GPa. The region including the first minimum is enlarged in the inset. Each RDF result is averaged over a 5-ps trajectory (50 successive snapshots with a time-interval of 0.1 ps) at the given pressure point during the compression.

## Supplementary References